\documentclass[preprint,aps,prb]{revtex4}
\usepackage{graphicx}
\usepackage{dcolumn}
\usepackage{bm}

\usepackage{titlesec}
\titleformat*{\section}{\flushleft \bf \large}
\titleformat*{\subsection}{\flushleft \bf}
\titleformat*{\subsubsection}{\flushleft}
\begin{document}

\title{Rotation of hydrogen molecules during the dissociative adsorption on
the Mg(0001) surface: A first-principles study}

\author{Yanfang Li$^{1,2}$, Yu Yang$^{2,3}$\footnote{Corresponding
author. E-mail address: yang\underline{ }yu@iapcm.ac.cn}, Bo
Sun$^2$, Hong-Zhou Song$^2$, Yinghui Wei$^1$, Ping
Zhang$^2$\footnote{Corresponding author. E-mail address:
zhang\underline{ }ping@iapcm.ac.cn}}

\affiliation{$^1$College of Materials Science and Engineering,
Taiyuan University of Technology,Taiyuan 030024, People's Republic
of China}

\affiliation{$^2$LCP, Institute of Applied Physics and Computational
Mathematics, P.O. Box 8009,Beijing 100088, People's Republic of
China}

\affiliation{$^3$Center for Advanced Study and Department of
Physics, Tsinghua University, Beijing 100084, People's Republic of
China}

\date{\today}

\begin{abstract}
Using first-principles calculations, we systematically study the
potential energy surfaces and dissociation processes of the hydrogen
molecule on the Mg(0001) surface. It is found that during the
dissociative adsorption process with the minimum energy barrier, the
hydrogen molecule firstly orients perpendicular, and then rotates to
be parallel to the surface. It is also found that the orientation of
the hydrogen molecule at the transition state is neither
perpendicular nor parallel to the surface. Most importantly, we find
that the rotation causes a reduction of the calculated dissociation
energy barrier for the hydrogen molecule. The underlying electronic
reasons for the rotation of the hydrogen molecule is also discussed
in our paper.
\end{abstract}

\maketitle

\section{Introduction}

It is of great scientific importance to study the dissociation of
diatomic molecules on metal surfaces (dissociative sticking), to
meet some intrinsic interests on mechanisms for bond breaking and
bond formation and origins of activation energy barriers
\cite{Darling1995,King1988}. The dissociation of the hydrogen
molecule (H$_{2}$) on metal surfaces are furthermore key events for
lots of technological applications such as hydrogen storage for
fuels \cite{Jacobson2002,Wu2008}, hydrogen caused embrittlement
\cite{Lu2005}, and heterogeneous catalysis
\cite{King1988,Gustafsson2006}. Among all the metals, the Mg(0001)
surface is one of the most studied prototypes, both theoretically
and experimentally
\cite{Sprunger1991,Norskov1981,Bird1993,Vegge2004,Wu2008,Johansson2006}.
However, some basic problems concerning the dissociation process and
energy barrier still have discrepancies yet. As early as 1981, N{\o
}rskov \textit{et al.} for the first time studied the dissociation
of H$_{2}$ on the Mg(0001) surface, observing a molecularly adsorbed
precursor state and an energy barrier of 0.5 eV for dissociation
\cite{Norskov1981}. Later, by using the more precise
first-principles methods, Bird \textit{et al.} discovered that
although the most energetically favorable site for dissociation is
the bridge site as reported in N{\o}rskov's paper, the dissociation
energy barrier is 0.37 eV, rather than the reported 0.5 eV, and
there is no precursor state \cite{Bird1993}. However, the
discrepancy on the dissociation energy barrier remains ever since
then. Vegge \textit{et al.} reported a value of 1.15 eV for the
dissociation, after systematically considered the zero point energy
of H$_{2}$ within their first-principles calculations
\cite{Vegge2004}. Wu \textit{et al.} reported a value of 1.05 eV by
employing first-principles calculations and transition state theory
\cite{Wu2008}. Meanwhile, Johansson \textit{et al.} recently
organized an experiment to study the dissociation energy barrier for
H$_2$ at Mg(0001), and the obtained barrier is 0.6 $\sim$ 0.9 eV
\cite{Johansson2006}, which disagrees with all the reported
theoretical values. So it is still an open question as to the
correct dissociation energy barrier for H$_{2}$ at Mg(0001).

On the other hand, it has been recently suggested that steric
effects might be important during the adsorption and dissociation of
diatomic molecules on metal surfaces \cite{Hou1997}. For the
adsorption of D$_{2}$ on Cu(111), both experimental and theoretical
investigations found that dissociation of D$_{2}$ occurs
preferentially when the molecule approaches with its bond parallel
to the surface \cite{Hou1997}. Similar dependence on polar angle has
also been theoretically predicted for the dissociative sticking of
O$_{2}$ on the Al(111) surface \cite{Osterlund1997}. When
considering the dissociative adsorption of H$_{2}$ on metal
surfaces, steric effects might be more important because of the low
inertia moment of H$_{2}$ and correspondingly high possibility to
rotate. In fact, it has already been pointed out that during the
dissociative adsorption of H$_{2}$ on the Pd(111) surface, the
molecular axis orientation has a drastic effect and low activation
barriers are only met over a small range of $\theta$ values from
parallel the surface \cite{Busnengo2000}. And for the adsorption on
the NiAl(110) surface, H$_{2}$ molecules rotate abruptly when they
are close to the surface, which allows them to adopt the orientation
that is more convenient for dissociation (i.e., nearly parallel to
the surface) \cite{Riviere2006}. For the adsorption of H$_{2}$ on
the Mg(0001) surface, however, this important issue of steric
effects has not been considered yet. Motivated by this observation,
here by using the first-principles calculations, we study the
rotation of H$_{2}$ during the dissociative adsorption on the
Mg(0001) surface and its corresponding influences on the
dissociation energy barrier. We show that the most energetically
favorable path for H$_{2}$ dissociation at Mg(0001) is fundamentally
determined by the steric effect.

\section{Calculation method}

Our calculations were performed within DFT using the Vienna
\textit{ab-initio} simulation package \cite{VASP}. The PW91
\cite{PW91} generalized gradient approximation and the
projector-augmented wave potential \cite{PAW} were employed to
describe the exchange-correlation energy and the electron-ion
interaction, respectively. The cutoff energy for the plane wave
expansion was set to 250 eV. The Mg(0001) surface was modeled by a
five-atomic-layer slab plus a vacuum region of 20 \AA. A $2\times2$
supercell was adopted in the study of the H$_{2}$ adsorption since
our test calculations have showed that it is large enough to avoid
the interaction between adjacent hydrogen molecules. Integration
over the Brillouin zone was done using the Monkhorst-Pack scheme
\cite{Monkhorst} with $9\times9\times1$ grid points. The calculated
lattice constants of bulk Mg ($a$, $c$) and the bond length of
isolated H$_{2}$ are 3.207 \AA, 5.145 \AA~ and 0.748 \AA,
respectively, in good agreement with the experimental values of 3.21
\AA, 1.62 \AA~ \cite{Amonenko1962} and 0.74 \AA~ \cite{Huber1979}.
The calculation of the potential energy surface was interpolated to
209 points with different bond length ($d_{\mathrm{H-H}}$) and
height ($h_{\rm H_2}$) of H$_{2}$ at each surface site.

\section{Results and discussion}

After geometry optimization for the Mg(0001) surface, we build our
model to study the potential energy surface (PES) of H$_{2}$ on the
relaxed Mg surface. As shown in Fig. 1, there are four different
high-symmetry sites on the Mg(0001) surface, respectively the top,
bridge (bri), hcp and fcc hollow sites. After PES calculations, we
find that the dissociation barrier of H$_{2}$ at low-symmetry sites
is always larger than at high-symmetry sites, proving that
high-symmetry sites play crucial roles in the adsorption of diatomic
molecules, similar to that has been observed on the adsorption of
oxygen molecules on the Pb(111) surface \cite{Yang2008}. And in the
following, we will only give the results at the four high-symmetry
sites. At each surface site, an adsorbed H$_{2}$ has three different
principle orientations, respectively along the $x$ (i.e.,
[$11\bar{2}0$]), $y$ (i.e., [$\bar{1}100$]), and $z$ (i.e.,
[$0001$]) directions. Herein, we use top-$x,y,z$, bri-$x,y,z$,
hcp-$x,y,z$ and fcc-$x,y,z$ respectively to represent the total
twelve high-symmetry channels for the adsorption of H$_{2}$ on the
Mg surface.

Throughout our PES calculations, we find no molecular adsorption
precursor states for H$_{2}$ at Mg(0001), according well with all
previous reports, except for the one by N{\o }rskov \textit{et al.}
using jellium model \cite{Norskov1981}. Our calculated result for
the lowest dissociation energy barrier, as well as that in other
theoretical and experimental reports, is given in Table. I. The
minimum energy path (MEP) for the dissociation of H$_{2}$ on the
Mg(0001) surface is found to be along the bri-$y$ channel, which is
consistent with all previous first-principles studies. The
transition state obtained from our PES calculations along the
bri-$y$ channel is at the point where $d_{\mathrm{H-H}}$=1.12 \AA~
and $h_{\rm H_2}$=1.16 \AA, which accords with previous results
\cite{Du2005,Wu2008}. However, as we will see later, This transition
state needs to be modified after considering the rotational degree
of freedom of H$_{2}$.

At the bridge site, however, we find that the total energy of the
H$_{2}$/Mg system is not always smaller along the bri-$y$ channel
than along other channels. This is a key point in this paper. In
fact, at large values of H$_{2}$ height from Mg(0001) surface, we
find that the total energy is smaller along the bri-$z$ channel than
along the bri-$y$ channel. To show this, we plot in Figs. 2(a)-(c)
the two-dimensional cuts of the PES along the bri-$x,y,z$ channels.
Correspondingly, the minimum energy paths in Figs. 2(a)-(c) are
collected and plotted in Fig. 2(d). It can be clearly seen that
there is a prominent crossing point in the minimum energy paths
along bri-$y$ and bri-$z$ channels, at which the distance of the
H$_{2}$ molecule from the surface takes a value of
$h_{\mathrm{H_{2}}}$=1.26 \AA. Before this crossing, the total
energy of the adsorption system along the bri-$z$ channel is always
lower than along the bri-$y$ channel. This finding indicates that
H$_{2}$ prefers to orient perpendicular to the Mg(0001) surface
until it reaches the height lower than 1.26 \AA. After the crossing
point, whereas, the system along the bri-$y$ channel has a smaller
total energy than along the bri-$z$ channel, and H$_{2}$ tends to
rotate from the bri-$z$ channel to the bri-$y$ channel.

We then further study the influence of the molecular rotation on the
dissociation energy barrier of H$_{2}$. For this we have calculated
the total energy by fixing the mass center of H$_{2}$ at the bridge
site with the height of 1.16 \AA~ and the molecular bond length of
1.12 \AA, while allowing H$_{2}$ to rotate around its mass center in
the $y$-$z$ plane (see the inset in Fig. 3). The calculated angle
dependence of the total energy is shown in Fig. 3. Clearly, it can
be seen that the transition state (namely, the saddle point in the
PES for H$_{2}$ dissociation) should be the structure where the
orientation of H$_{2}$ is 21$^{\circ}$ from the $x$-$z$ plane. This
rotation of H$_{2}$ results in a 86 meV modification on the
dissociation energy barrier. This finding suggests that steric
factors that has not been considered in previous theoretical
calculations might be (at least partially) responsible for their
discrepancies with experimental measurement.

Although it has long been explored for steric effects on the
dissociative adsorption of diatomic molecules on metal surfaces, the
specific reasoning has seldom been discussed yet. Herein we will try
to find the underlying mechanisms on the rotation of H$_{2}$ during
the dissociative adsorption on the Mg(0001) surface, by analyzing
carefully the charge distributions and electronic interactions along
the adsorption process of H$_2$. Figures 4(a) and (b) show the
difference electron density for the adsorption system with $h_{\rm
H_2}$ to be 2.00 and 1.16 \AA~ along the bri-$y$ channel, namely,
\begin{equation}\label{Drho}
\Delta\rho =\rho({\rm H_2+ Mg(0001)}) - \rho({\rm H_2}) - \rho({\rm
Mg(0001)})],
\end{equation}
where $\rho({\rm H_2+ Mg(0001)})$, $\rho({\rm H_2})$ and $\rho({\rm
Mg(0001)})$ are respectively the electron density of the adsorption
system, the H$_2$ molecule and the clean Mg(0001) surface. To
calculate $\Delta\rho$, the atomic positions in the last two terms
in Eq. 1 have been kept at those of the first term. Through careful
wavefunction analysis, we find that at the beginning of the
adsorption process, the molecular orbitals of H$_2$ orthogonalize
with electronic states of Mg and thus are broadened. As shown in
Fig. 4(a), the surface electrons of Mg are repelled from the region
occupied by the H$_2$ bonding electrons due to these
orthogonalizations. This interaction has also been observed during
the interacations of H$_2$ with other metals such as the Al(111)
\cite{Hammer1993} and transition metal surfaces \cite{Harris1985}.
When the H$_2$ molecule is close enough to the Mg(0001) surface and
come to the transition state for its dissociation, electrons
transfer from electronic states of Mg to the antibonding orbital of
H$_2$, which can be clearly seen from Fig. 4(b). We can also see
from Fig. 4(b) that the orthogonalizations between electronic states
of Mg and the bonding orbital of H$_2$ still exist at the transition
state.

In total, the orthogonalizations between molecular orbitals of H$_2$
and electronic states of Mg cause repulsive interactions between
electrons of H and Mg, and thus will enlarge the total energy of the
adsorption system, while the electrons transfer from Mg to H$_2$
causes attractive interactions between H and Mg atoms and lowers
down the total energy. So during the adsorption process of H$_2$,
the total energy of the system firstly goes up, then lowers down.
This analysis explains why an energy barrier is needed for the
dissociation of H$_2$ on the Mg(0001) surface. Moreover, both the
orthogonalizations and electrons transfer are always weaker along
the bri-$z$ channel than along the bri-$y$ channel. Therefore, at
the beginning of the adsorption process, when no electrons transfer
happens, the total energy of the system is smaller along the bri-$z$
channel than along the bri-$y$ channel. And at around the transition
state, when the H$_2$ molecule is very close to the Mg(0001)
surface, electrons transfer begins to dominate the molecule-metal
interaction. So the total energy along the bri-$y$ channel is
smaller at the transition state. Herein, the rotation of H$_2$ can
be seen as the result from the different interactions that
respectively favors the bri-$y$ and bri-$z$ channels. As a result,
the minimum energy path for the dissociation of H$_2$ is neither
along the bri-$y$ nor along the bri-$z$ channels. And the
corresponding transition state is the one where H$_{2}$ orients
21$^{\circ}$ away from the $x$-$z$ plane, as shown in Fig. 3.

\section{Conclusion}

In conclusion, we have systematically studied the PESs for the
dissociative adsorption of the hydrogen molecule on the Mg(0001)
surface. Our results accord well with previous reports on the direct
dissociative adsorption process. More importantly, we have found
that the hydrogen molecule does not always orient parallel to the
surface along the dissociation channel with the lowest energy
barrier. At large molecular heights, H$_{2}$ orients perpendicular
to the surface. When getting closer to the surface, H$_{2}$ begins
to rotate such that at the transition state, H$_{2}$ orients
21$^{\circ}$ away from the $x$-$z$ plane, which causes a $86$ meV
modification on the dissociation energy barrier. We have revealed
that this molecular rotation is because of the two different
interactions between H and Mg, i.e., the orthogonalizations between
molecular orbitals of H$_2$ and electronic states of Mg and
electrons transfer from the Mg(0001) surface to the antibonding
orbital of H$_2$. As a final concluding remark, here based on the
present results, we would like to point out that steric effects are
important to understand the adsorption behaviors of H$_{2}$ on metal
surfaces.

\begin{acknowledgments}
P.Z. was supported by the NSFC under Grants No. 10604010 and No. 60776063.
Y.W. was supported by the NSFC under Grants No. 50471070 and No. 50644041.
\end{acknowledgments}

\clearpage
\begin{table}[ptb]
\caption{ Our calculated minimum energy barrier for the dissociation
of H$_{2}$ molecules on the Mg(0001) surface, and other reported
results.}
\centering%
\begin{tabular}
[c]{ccc}\hline\hline
references & Methods & dissociation barrier (eV)\\
\onlinecite{Norskov1981} & Jellium model (LDA) & 0.50\\
\onlinecite{Bird1993} & DFT (LDA) & 0.37\\
\onlinecite{Jacobson2002} & DFT (RPBE) & 0.50\\
\onlinecite{Vegge2004} & DFT (GGA) & 1.15\\
\onlinecite{Du2005} & DFT (LDA) & 0.35\\
\onlinecite{Du2005} & DFT (PBE-GGA) & 1.05\\
\onlinecite{Wu2008} & DFT (GGA) & 1.05\\
\onlinecite{Johansson2006} & Experiment & 0.59$\sim$0.90\\
this work & DFT (GGA) & 0.85\\\hline\hline
\end{tabular}\label{barrier}
\end{table}
\clearpage

\noindent\textbf{List of captions} \\

\noindent\textbf{Fig.1}~~~ (Color online). (a) The $p$($2\times2$)
surface cell of Mg(0001) and four on-surface adsorption sites. Here
only the outmost two layers of the surface are shown. (b) The sketch
map showing that the molecule (with vertical or parallel
orientation) is initially away from the surface with a hight
$h_{H_2}$. \\

\noindent\textbf{Fig.2}~~~ (Color online). Contour plots of the two
dimensional cuts of the potential energy surfaces (PESs) for the
H$_{2}$/Mg(0001) system as a function of the bond lengths
($d_{\mathrm{H-H}}$) and the heights ($h_{\mathrm{H_{2}}}$), with
H$_{2}$ at the bridge site orienting along $x$ (a), $y$ (b) and $z$
(c) directions. (d) Minimum energy paths obtained along the three
different channels with different heights of H$_{2}$
($h_{\mathrm{H_{2}}}$) from the Mg surface. \\

\noindent\textbf{Fig.3}~~~ (Color online). The total energy of the
adsorption system with different orientations of H$_{2}$ at the
transition state point along the bri-$y$ channel where
$d_{\mathrm{H-H}}$=1.12 \r{A}~ and $h_{\mathrm{H_{2}}}$=1.16 \r{A}.
The inset depicts the definition of the angle $\alpha$, in which the
grey area and blue balls respectively represent the Mg surface and
two H atoms. \\

\noindent\textbf{Fig.4}~~~  (Color online). The difference electron
density for the H$_2$/Mg(0001) system with the height of H$_2$ to be
2.00 (a) and 1.16 \AA~ (b). Blue and grey balls respectively
represent hydrogen and Mg atoms. Dark and dashed lines respectively
represent plus and negative values, i.e., electrons accumulation and
depletion.\\

\clearpage

\begin{figure}
\includegraphics[width=1.0\textwidth]{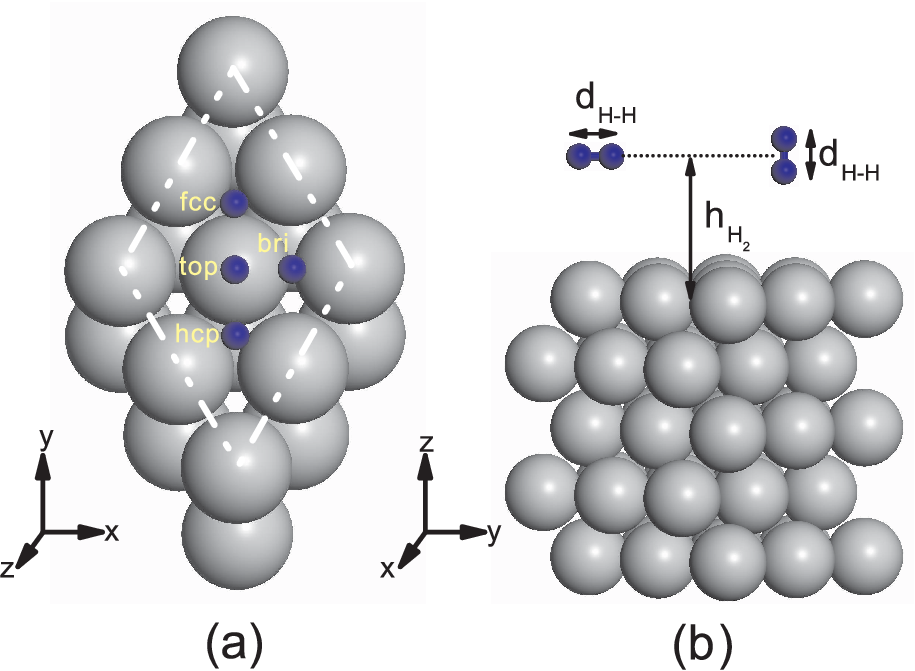}
\caption{\label{fig:fig1}}
\end{figure}
\clearpage
\begin{figure}
\includegraphics[width=1.0\textwidth]{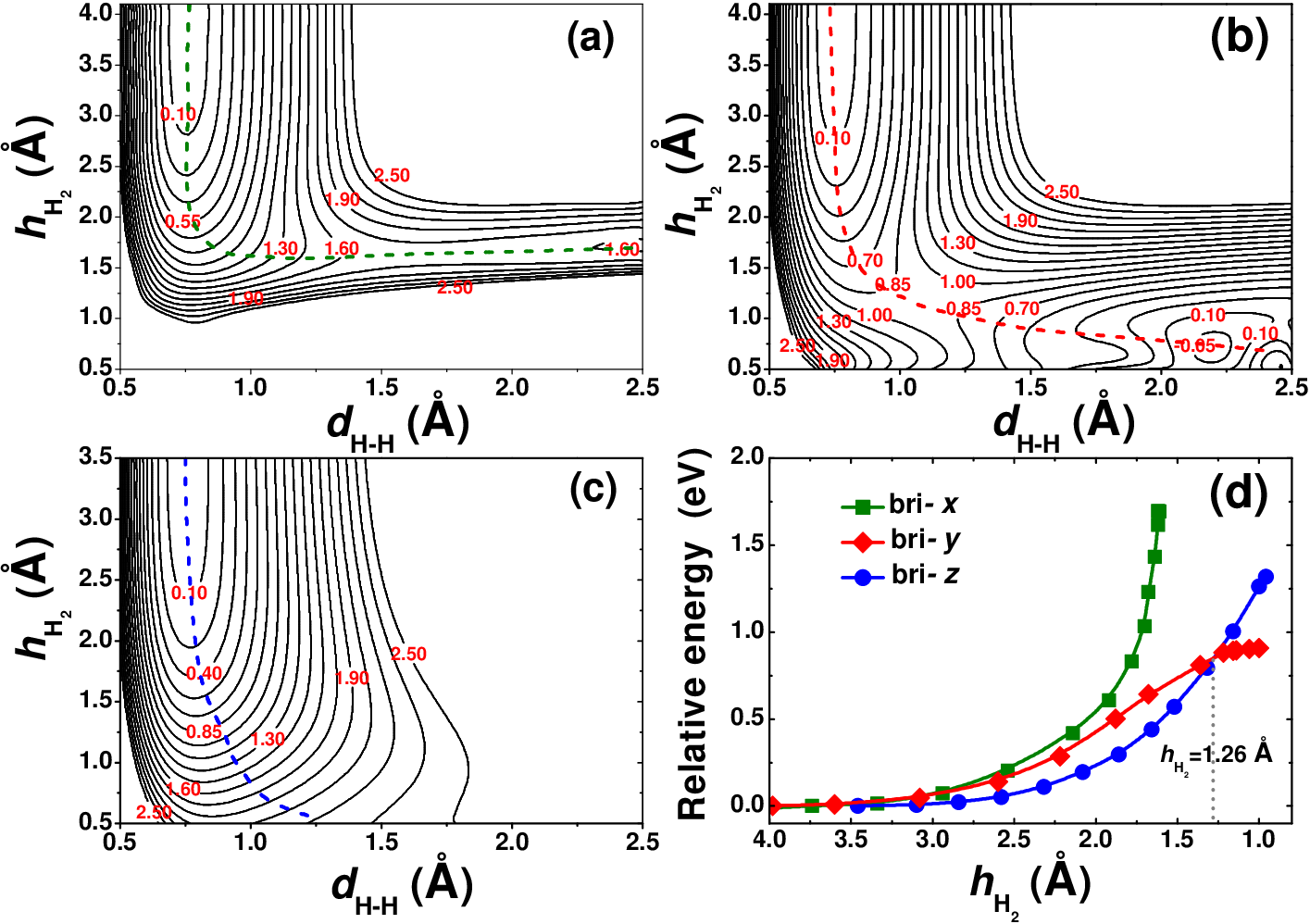}
\caption{\label{fig:fig2}}
\end{figure}
\clearpage
\begin{figure}
\includegraphics[width=1.0\textwidth]{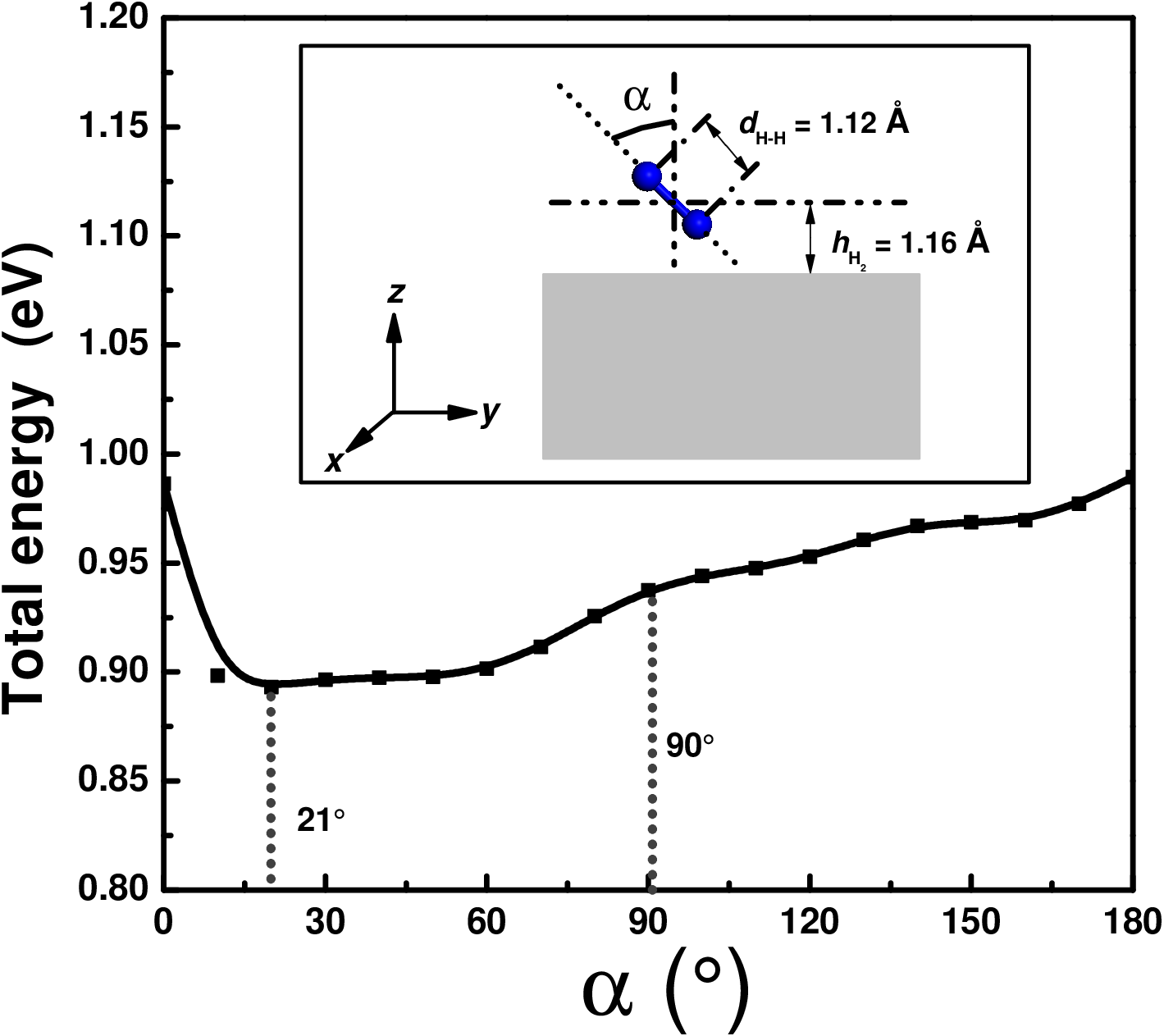}
\caption{\label{fig:fig3}}
\end{figure}
\clearpage
\begin{figure}
\includegraphics[width=1.0\textwidth]{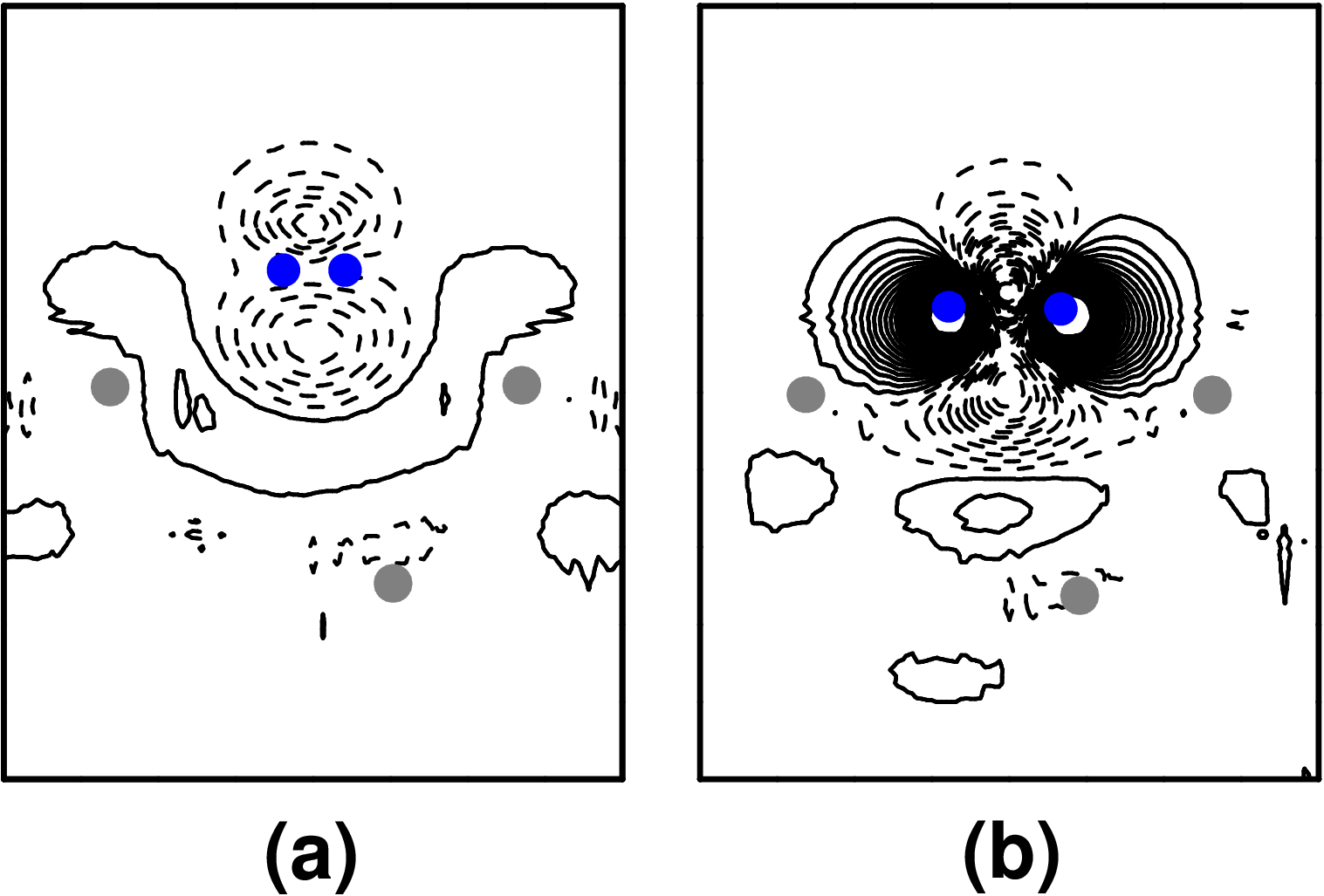}
\caption{\label{fig:fig4}}
\end{figure}
\clearpage
\end{document}